\documentclass[3p,times,authoryear]{elsarticle}

\usepackage{lineno,hyperref}
\usepackage{graphicx}
\usepackage{amsmath}
\usepackage{amsfonts}
\usepackage{amssymb}
\usepackage{url}
\usepackage[T1]{fontenc}
\usepackage{mathptmx}      
\usepackage{multirow}
\usepackage{lscape}
\modulolinenumbers[1]
\journal{}

\begin{document}
\begin{frontmatter}
\title{Quantification of historical drought conditions over different climatic zones of Nigeria}

\author[mymainaddress]{Samuel T. OGUNJO \corref{mycorrespondingauthor}}
\ead{stogunjo@futa.edu.ng, +2348026009902}
\author[mymainaddress]{Oluwatobi O. Ife-Adediran}
\author[mymainaddress]{Eunice O. Owoola}
\author[mymainaddress,mysecondaryaddress]{Ibiyinka A. FUWAPE}

\address[mymainaddress]{Department of Physics, Federal University of Technology, Akure, PMB 704 Akure, Ondo State, Nigeria}
\address[mysecondaryaddress]{Michael and Cecilia Ibru University, Ughelli North, Delta State, Nigeria.}

\begin{abstract}
The impact of extreme climate such as drought and flooding on agriculture, tourism, migration and peace in Nigeria is immense.  There is the need to study the trend and statistics for better planning, preparation and adaptation.  In this study, the statistical and temporal variation of climatic indices Standardized Precipitation Index (SPI ) and Standardized Precipitation Evapotranspiration Index (SPEI) were computed for eighteen (18) stations covering four climatic zones (Sahel, Midland, Guinea Savannah and Coastal) of tropical Nigeria.  Precipitation, minimum and maximum temperature from 1980 - 2010  obtained from the archives of the Nigerian Meteorological Services were used to compute both the SPI and SPEI  indices at 1-, 3- 6- and 12-month timescales.  The temporal variation of drought indices showed that droughts were more prominent at 6- and 12-months timescales.  SPI and SPEI were found to be better correlated at longer timescales than short time scales.  Predominant small, positive and significant trend across the region suggest an increasing trend due to climate change.
\end{abstract}

\begin{keyword}
  climate indices \sep Standardized Precipitation Index \sep Standardized Precipitation Evapotranspiration Index \sep Nigeria \sep climate change.
\end{keyword}
\end{frontmatter}
\newpage

\section{Introduction}
Drought is an unusual period of dryness as a result of low precipitation or high temperature \citep{sordo2017analysis}. A drought event is characterized by a continuous shortage of water due to low rainfall over a period of time  \citep{chen2009historical}. The occurrence of drought in a location over a long period of time is referred to as severe drought (Muhammad et al. 2017). Droughts have become a consistent global climatic occurrence \citep{pereira2009coping}. Intense drought conditions have been linked to the accumulation of greenhouse gases especially from decades of industrial activities \citep{gudmundsson2015european,stocker2014climate,field2012managing}. Drought events in different regions of the globe vary in intensity (severity or magnitude), frequency of occurrence and duration \citep{wilhite1993planning,dracup1980}.

The effects of drought are far reaching to agriculture, ecology, health. Droughts that have direct effects on crop growth and yield as a result of dryness in their roots is referred to as agricultural droughts. \citet{ahmad2004drought} reported that agricultural drought occurred in Pakistan during years 2000 and 2001.  During the same period, there was a severe drought in North Korea that led to a significant drop in food production \citep{josserand2008fao}. Other types of physical droughts are: meteorological and hydrological droughts \citep{khan2018detecting}.  \citet{keyantash2002quantification} described a non-physical form of drought referred to as socioeconomic drought. \citet{blain2012revisiting} emphasized the slowly accumulating effect of drought and the need for its early detection. Notably, there is an expected increase in the severity, spread and effect of drought in some African countries by 2020. \citep{pachauri2008changements}.

Drying trends were found to be increasing from the 1970s till the 1990s in a study carried out by \citet{qian2001climate} for regions of Northern China.  \citet{Oloruntade2017} found high correlations values of 0.65 and 0.55 between SPI and SPEI at 3-months and 12-months in the Niger-South Basin area of Nigeria.  SPI and SPEI values were found to produce similar droughts characteristics over the Volta basin with a correlation of 0.97 in observed and simulated data \citep{oguntunde}.  Drought can lead to violence and radical attitudes among politically neglected, marginalized and disaffected people \citep{Detges2017}.   Positive correlation has been reported between Atlantic Nino 1 along the coastal regions of West Africa while a negative correlation prevails in the Sahel region \citep{Adeniyi2018}.  Coupled ocean-atmosphere phenomena has been found to influence drought events within the Greater Horn of Africa region \citep{Mpelasoka2018}.  Drought vulnerability index over Africa during the period 1960 - 2100 showed that northern African countries such as Egypt, Tunisia, and Algeria are the least drought vulnerable countries \citep{AHMADALIPOUR2018520}.  \citet{NDEHEDEHE2016106} reported that El-Nino Southern Oscillation (ENSO), Atlantic multi-decadal oscillation and Atltantic meridional mode (AMM) are associated with extreme rainfall conditions with statistically significant relationship between AMO and SPI at 12 months.

Studies on drought trends are useful in the development of models for future hydroclimatic predictions and water resource planning. A number of drought assessment indices are existent and in use \citep{khan2018detecting}.  The choice of a drought index depends on the type of drought as each index in use has its uniqueness and application some of which were highlighted in \citep{khan2018detecting} and \citet{mishra2010review}. Indicators such as the Standard Precipitation Index (SPI) can be used to characterize droughts \citep{moreira2008spi}. SPI has an advantage in its usefulness for impact assessment of agricultural droughts \citep{zargar2011review}. However the SPI is also recommended for the characterization of meteorological droughts all over the world \citep{blain2012revisiting}.  The Standard Precipitation Evapotranspiration Index (SPEI) factors evapotranspiration and precipitation in its computation. Other indicators that have been popularly used in the characterization of drought events include: Reconnaissance Drought Index (EDI), Palmer Drought Severity Index (PDSI), Surface Water Supply Index (SWSI) and Effective Drought Index (EDI).

Studies on drought over Nigeria have always considered subregions. In this study, we aim to investigate the statistics of drought across the different climatic zones of Nigeria using SPI and SPEI.  The correlation between the two drought indices will be investigated, as well as, their regression analysis, trend and frequency distributions.  Results from this study is expected to show the comparative drought risks across the different regions in Nigeria and provide information for strategic planning and adaptation.

\section{Methodology}\label{methods}
\begin{figure}[!h]
\centering
  \includegraphics[scale=0.5]{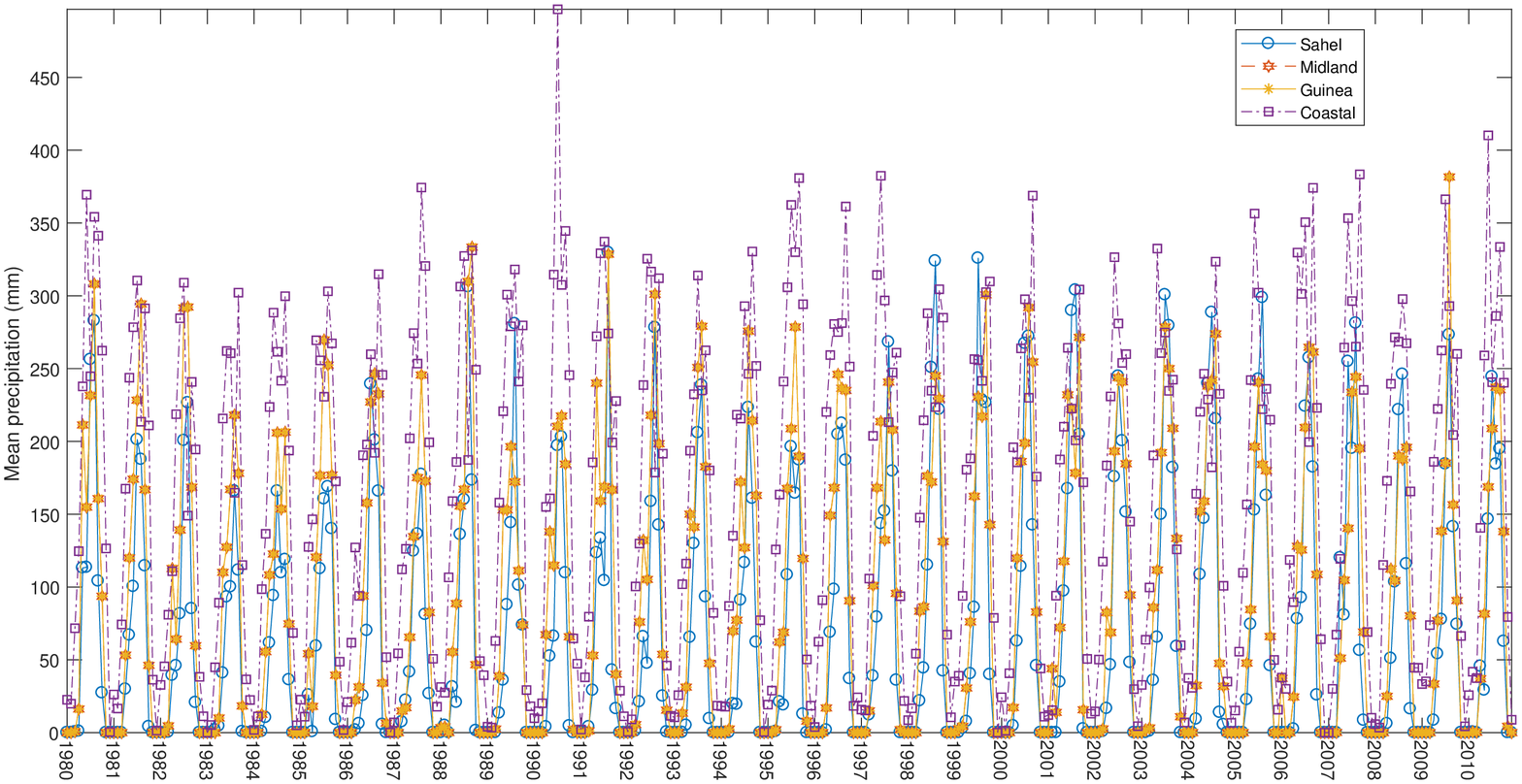}\\
  \caption{Mean monthly precipitation across the regions.}\label{fig1}
\end{figure}

Eighteen locations across the four climatic zones of Nigeria were considered in this study.  The geographical coordinates and statistics of the location are presented in Table \ref{tab1} and the temporal variation of mean regional precipitation is shown in Figure \ref{fig1}.   Monthly precipitation, minimum temperature and maximum temperature data were obtained from the archives of the Nigerian Meteorological Services from 1980 - 2010.

The Standardized Precipitation Index (SPI) is based on the use of probability distribution to evaluate the departure of precipitation during a time span from the mean value \citep{oguntunde}.  The parameters, $\alpha$ and $\beta$, of a gamma distribution obtained are from maximum likelihood estimation when precipitation data is fitted to a gamma distribution given by

\begin{equation}\label{gammsa}
G(x) = \frac{1}{\beta^\theta\Gamma(\theta)}\int_0^x x^{\theta -1}e^{-x/\beta}dx
\end{equation}
where $x>0$, $\Gamma$ is the gamma function, $\alpha$ and $\beta$ are the form and scale parameter respectively.  The Standardized Potential Evaporation Index is computed in the same way, however, with the consideration of potential evapotranspiration \citep{oguntunde,Oloruntade2017}.  The drought indices are classified as

\begin{equation}
SPI = \left\{
  \begin{array}{ll}
    SPI \geq 2.00, & \hbox{Extremely wet (EW);} \\
    1.50\leq SPI < 2.00, & \hbox{Very wet (VW);} \\
    1.00\leq SPI < 1.50, & \hbox{Moderately wet (MW);} \\
    -1.00\leq SPI < 1.00, & \hbox{Near Normal (NN);} \\
    -1.50\leq SPI < -1.00, & \hbox{Moderately drought (MD);} \\
    -2.00\leq SPI < -1.50, & \hbox{Severely drought (SD);} \\
    SPI < -2.00, & \hbox{Extreme drought (ED).}
  \end{array}
\right.
\end{equation}
\begin{table}
  \centering
  \caption{Geographical coordinates and statistics of study locations in the period 1981 - 2010.}\label{tab1}
  \begin{tabular}{|l|lllrrr|}
    \hline
Region	&Location	&LATITUDE	&LONGITUDE	&TOTAL (mm)	&MEAN (mm)	&STDEV\\ \hline
\multirow{4}{*}{Sahel}	&Sokoto	&13.03	&5.12	&36057.0	&96.93	&116.29\\
	   &Maiduguri	&11.5	&13.1	&17847.7	&47.98	&75.60\\
	   &Kano	&12.02	&8.32	&31236.3	&83.97	&130.51\\
	   &Katsina	&12.51	&7.33	&15390.3	&41.37	&83.50\\ \hline
\multirow{3}{*}{Midland}	&Kaduna	&10.28	&7.25	&32226.0	&86.63	&112.66\\
	   &Minna	&9.38	&6.31	&37312.2	&100.30	&106.65\\
	   &Yola	&9.15	&12.3	&27500.9	&73.93	&84.38\\ \hline
\multirow{3}{*}{Guinea Savannah}	&Lokoja	&7.47	&6.37	&38280.5	&102.90	&101.79\\
	&Markudi	&7.45	&8.53	&36528.8	&98.20	&109.41\\
	&Warri	&5.52	&5.75	&85674.7	&230.31	&189.02\\ \hline
\multirow{8}{*}{Coastal}	&Lagos	&6.61	&3.62	&50944.4	&136.95	&146.50\\
	&Akure	&7.25	&5.19	&44495.6	&119.61	&94.67\\
	&Port Harcourt	&4.41	&6.59	&71602.6	&192.48	&147.79\\
	&Owerri	 &5.27	&6.59	&73599.8	&197.85	&161.34\\
	&Enugu	&6.27	&7.29	&53822.4	&144.68	&131.01\\
	&Calabar	&4.58	&8.21	&90572.6	&243.47	&181.65\\
	&Ogoja	&6.66	&8.79	&58444.0	&157.11	&148.04\\
	&Abeokuta	&7.15	&5.15	&36002.0	&100.01	&90.81\\
    \hline
  \end{tabular}
\end{table}


\section{Results and Discussion}\label{results}

\begin{figure}[!h]
\centering
  \includegraphics[scale=0.5]{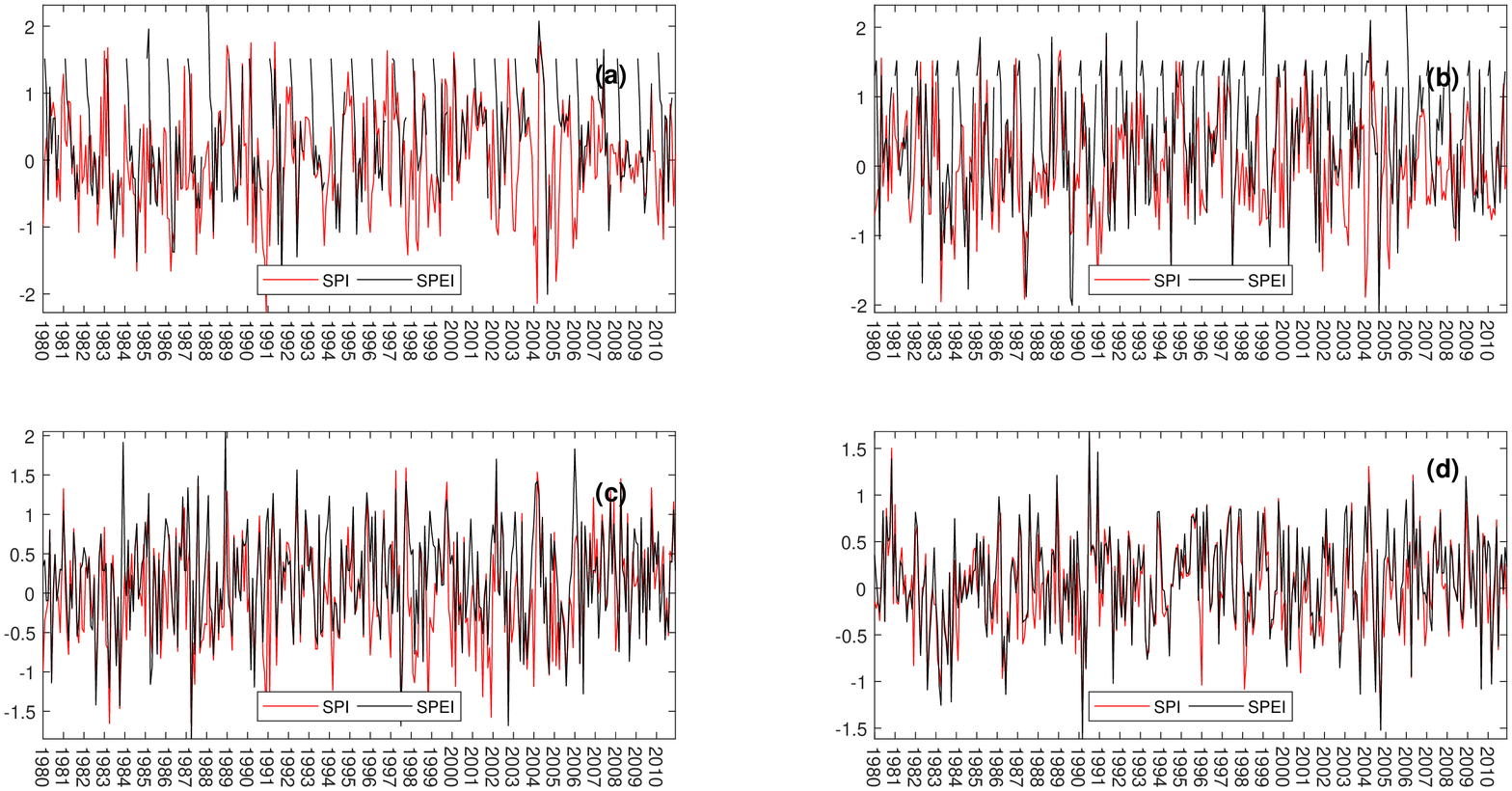}\\
  \caption{Mean SPI and SPEI values for (a)Sahel  (b) Midland (c) Guinea Savanna and (d) Coastal regions at 1-month time scale.}\label{fig2}
\end{figure}

\begin{figure}[!h]
\centering
  \includegraphics[scale=0.5]{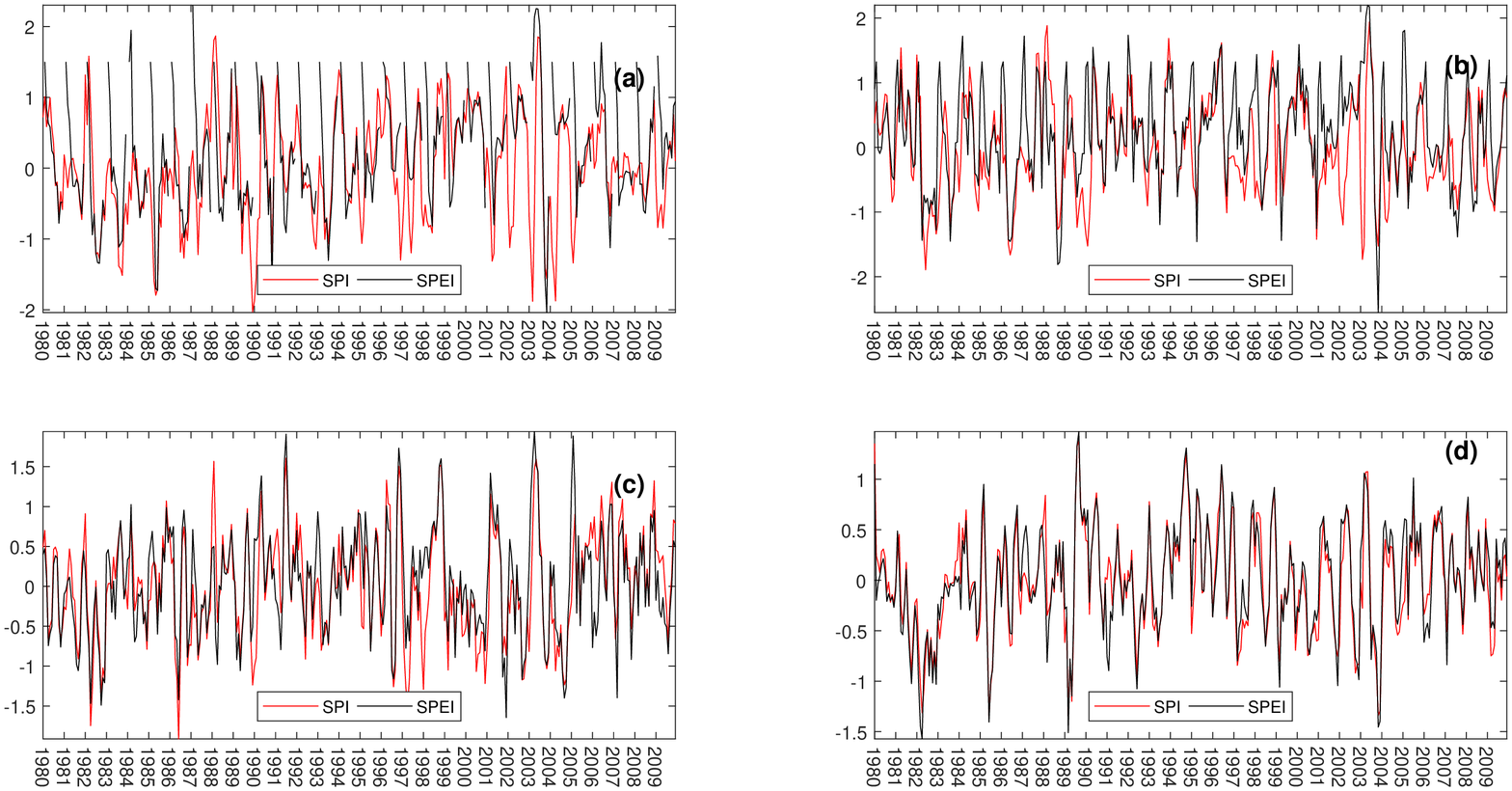}\\
  \caption{Mean SPI and SPEI values for (a)Sahel  (b) Midland (c) Guinea Savanna and (d) Coastal regions at 3-month time scale.}\label{fig3}
\end{figure}

\begin{figure}[!h]
\centering
  \includegraphics[scale=0.5]{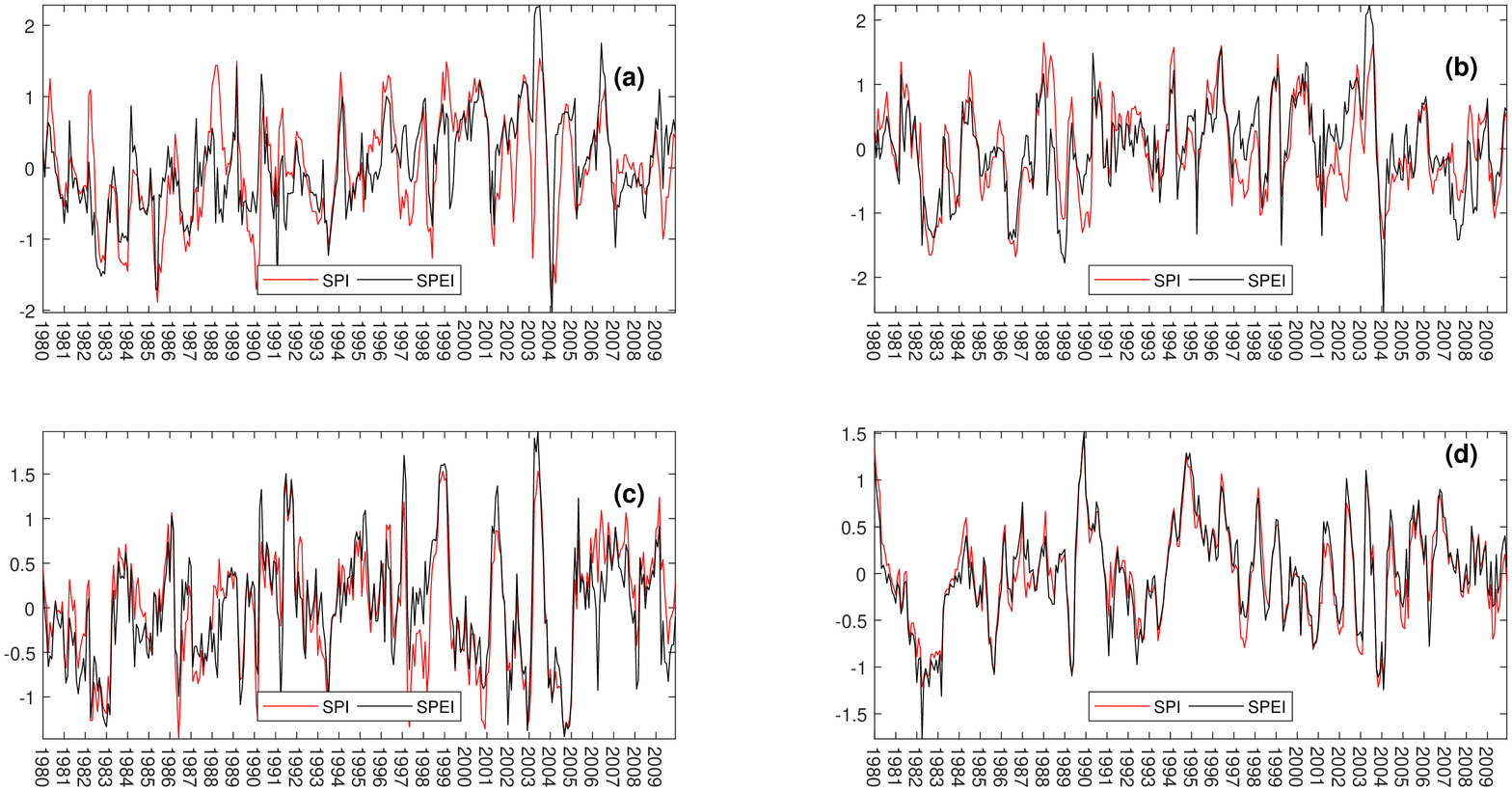}\\
  \caption{Mean SPI and SPEI values for (a)Sahel  (b) Midland (c) Guinea Savanna and (d) Coastal regions at 6-month time scale.}\label{fig4}
\end{figure}

\begin{figure}[!h]
\centering
  \includegraphics[scale=0.5]{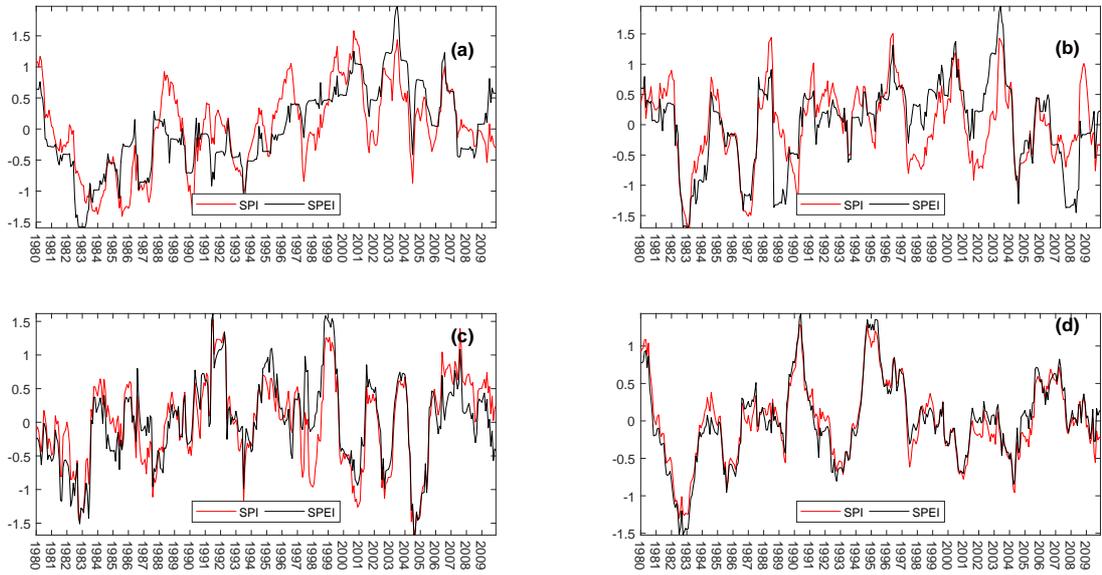}\\
  \caption{Mean SPI and SPEI values for (a)Sahel  (b) Midland (c) Guinea Savanna and (d) Coastal regions at 12-month time scale.}\label{fig5}
\end{figure}

The temporal evolution of mean SPI and SPEI values at 1-, 3-, 6- and 12-month timescales are presented in Figures \ref{fig2} - \ref{fig5} respectively.  The SPEI values at 1-month in the Sahel and Midland region showed periodic occurrences which can be attributed to the short but intense dry seasons experienced in the regions (Fig. \ref{fig2} a-b).  The drought intensity in these regions tend to be higher than the intensity in the Guinea Savannah and Coastal regions.   At 3-month scale (Figure \ref{fig3}), the seemingly periodicity in mean SPI/SPEI values persist in the Sahel but not in the Midlands.  A significant drought event could be seen in 1982-1983 in the Coastal region.  Considering the temporal variation of the indices at 6- and 12-month timescales, periodicity was no more obvious in the Sahel region.  Prominent drought regimes and general agreement between SPI and SPEI could be observed 1983, 1990, 1992/93 and 2004.  The drought period observed are in agreement with reported drought periods in the region  \citep{oguntunde,NDEHEDEHE2016106}.

The linear correlation between SPI and SPEI at different time scales (1,3,6,12) for the locations in each climatic zones is presented in Table \ref{corr_spi_spei}.   In the Sahel region, the strongest was observed at 12-month.  Maiduguri has the lowest correlation values at both 3- and 6-month scales while Sokoto and Katsina have the lowest correlation values between SPI and SPEI at 1- and 12-month scale respectively.  In the Midland area, Kaduna has the highest correlation values between SPI and SPEI while the lowest correlation values were observed in Yola.   Warri, in the Guinea savannah, was found to have the highest correlation values for the region at all scales.  The lowest correlation values were obtained at the 1-month scale.  An unusually low correlation value was observed at 3-month scale for Calabar in the coastal region.   The correlation values were observed to increase in the progression Sahel - Midland area - Guinea Savanna - coastal area.   This increasing trend could be attributed to the reducing effect of evapotranspiration as one move from the northern region to the coastal area of the country. In a study over the Niger-South basin of Nigeria, \citep{Oloruntade2017} obtained correlation in the range $0.56 - 0.66$.  This range is lower than the values obtained in this work for the same region.  The differences can be attributed to different data sources and time range.  SPI and SPEI have better comparative performance at longer time scales than shorter durations.

A linear regression of the two indices showed the same trend as correlation analysis (Table \ref{reg}).  The best fit in the Sahel region was found in Kano at 12-month scale with slope and $r^2$ values close to 1.  The weakest fits were observed in Sokoto at 1- and 3-month scales.  The regression fit for the Sahel region showed better fit at 12-month scale compared to other scales.  In the Midland region, the worst regression fit could be found in Yola.  The location has the lowest slope and $r^2$ values for the region. The performance of the fit were better in the Guinea Savannah and the Coastal region.  Calabar showed perfect fits with good coefficient of determination for the Coastal region.  The Guinea Savannah and Coastal regions of Nigeria have longer wet seasons, short dry seasons and larger amount of precipitation than the Midland and Sahel region.

Trend over the study period was computed for SPI and SPEI (Table \ref{trend}) using Mann-Kendall algorithm.  Trends in SPI and SPEI at all timescales were in the range $-0.00349 - 0.00587$ and $-0.00135 - 0.007$ respectively.  The trend values obtained for SPI are less than the range of $-0.026 - 0.011$ obtained in Southern Portugal at 12-month timescale \citep{Costa2011}.  In the Sahel region, all locations exhibit positive trends at all timescales except Sokoto which has negative trends.  In the Midland region, Minna has positive trends at all timescales while the other two locations in the region showed negative trends.  In the Guinea Savannah, Warri is the exception with negative trend at all time scales while in the Coastal region, Lagos, Akure and Owerri are the stations with negative trends.  The varying trend signs within regions are attributed to local dynamics and topography.  Locations in all the region have the same trend sign in both SPI and SPEI except Markurdi, Warri and Akure at the 1-month timescale; Lagos, Oweri and Akure at 3-month timescale; Yola, Lagos and Owerri at 6-month timescale; and Owerri at 12-month timescale.  All these locations have negative signs in SPI but positive signs in SPEI computation.  The trends were found to be significant at 95\% confidence interval except locations in Sahel and Midland regions at 1- and 3- months timescale for SPEI and Sokoto, Minna and Yola at 6- and 12-months timescales.  The trend values are similar to those obtained in Cyprus by \citet{Katsanos2018} but lower than that reported by \citet{oguntunde} for the same region at 12-month scale.  In the report by \citet{oguntunde}, a longer period of time was used, different data source as well as a different method of computing the slope was used, and ecological zones were considered rather than climatic zones at only 12-month timescale.

Frequency distribution of SPI and SPEI classes are shown in Tables \ref{stats1} - \ref{stats12} at 1-, 3-, 6-, and 12-months timescales respectively.  In the Sahel and Midland regions, the occurrences of near normal events were greater in SPI than SPEI at 1- and 3-months timescales as opposed to the occurrences being greater in SPEI for both Guinea Savannah and Coastal regions. This trend was not noticeable at 6- and 12-months timescales.   This implies that both SPI and SPEI have comparative performance at higher timescales than shorter timescales.  It can be inferred that atmospheric-land interactions such as El Nino and Atlantic Nino 1 influence drought at short time scales.  Atlantic Nino 1 has been reported to have positive correlation with SPEI in the coastal regions of West Africa but negative correlation in the Sahel regions \citep{Adeniyi2018}.  It has also been posited that El-Nino Southern Oscillation-ENSO, Atlantic Multi-decadal Oscillation-AMO, and Atlantic Meridional Mode-AMM can be responsible for the results obtained 12-month timescale \citep{NDEHEDEHE2016106}.  Near normal condition was found to be predominant in all the analysis.  A noticeable trend in all climatic zones is the the higher number of occurrences of MD values in SPI than the MD values obtained under SPEI.

\section{Conclusion}
In this study, the performance and statistics of two drought indices were investigated over different climatic zones of Nigeria.  The study considered eighteen locations over a period of thirty-one years.   At 1- and 3- months scale, both SPI and SPEI are predominantly periodic in the Sahel and Midland regions indicating a harsh and severe dry season.   At longer time scales, SPI and SPEI have stronger correlation than at shorter time scales.  This trend was also reflected in the regression coefficient between the two indices.  The regression fits were better in the southern part of the country than the northern part of the country.  Comparative trend analysis of drought over Nigeria using SPI and SPEI were also carried out.  Results obtained indicate that small but positive significant trends were predominant over the region in the period under consideration.  Finally, the statistical distribution of both SPI and SPEI at different time scales reveal the prevalence of near normal conditions.  It is posited that coupled ocean-atmosphere is responsible for the statistical distribution across the different climatic regions.

There is the need for spectral analysis to determine the frequency of occurrence of drought in the region.  Furthermore, the role of ocean-atmosphere coupling such as ENSO in drought frequency over the region is worth of investigation.

\begin{table}[!h]
\centering
  \caption{Linear correlation between SPI and SPEI for different locations at different time scales.}\label{corr_spi_spei}
\begin{tabular}{|l|l|llll|}

\multicolumn{2}{c}{}             & \multicolumn{4}{c}{Scale}       \\ \hline
Region                      & Location      & 1     & 3     & 6     & 12    \\ \hline
\multirow{4}{*}{Sahel}      & Sokoto        & 0.607 & 0.631 & 0.734 & 0.787 \\
                            & Maiduguri     & 0.651 & 0.609 & 0.645 & 0.736 \\
                            & Kano          & 0.699 & 0.693 & 0.783 & 0.899 \\
                            & Katsina       & 0.712 & 0.649 & 0.667 & 0.733 \\ \hline
\multirow{3}{*}{Midland}                            & Kaduna        & 0.726 & 0.706 & 0.830 & 0.869 \\
                             & Minna         & 0.650 & 0.718 & 0.792 & 0.781 \\
                            & Yola          & 0.639 & 0.548 & 0.622 & 0.542 \\ \hline
\multirow{3}{*}{Guinea Savannah }                            & Lokoja        & 0.765 & 0.793 & 0.855 & 0.877 \\
                           & Markudi       & 0.685 & 0.791 & 0.876 & 0.922 \\
                            & Warri         & 0.932 & 0.955 & 0.958 & 0.946 \\ \hline
\multirow{8}{*}{Coastal}               & Lagos         & 0.899 & 0.928 & 0.944 & 0.956 \\
                            & Akure         & 0.860 & 0.847 & 0.876 & 0.858 \\
                             & Port Harcourt & 0.932 & 0.908 & 0.935 & 0.954 \\
                            & Owerri        & 0.915 & 0.924 & 0.939 & 0.939 \\
                            & Enugu         & 0.796 & 0.876 & 0.942 & 0.953 \\
                            & Calabar       & 0.945 & 0.069 & 0.974 & 0.973 \\
                            & Ogoja         & 0.822 & 0.870 & 0.936 & 0.949 \\
                            & Abeokuta      & 0.827 & 0.849 & 0.861 & 0.859  \\\hline
\end{tabular}
\end{table}

\begin{table}[]
\centering
  \caption{Linear regression between SPI and SPEI for different locations at different time scales.}\label{reg}
  \small
\begin{tabular}{|l|l|llll||llll|}
\multicolumn{2}{c}{}                   &  \multicolumn{4}{c}{Regression slope}     &        \multicolumn{4}{c}{$r^2$}          \\ \hline
\multicolumn{2}{c}{}                    &  \multicolumn{4}{c}{Scale}   &         \multicolumn{4}{c}{Scale}         \\ \hline
Region          & Location           & 1     & 3     & 6     & 12    & 1        & 3        & 6        & 12       \\ \hline
\multirow{4}{*}{Sahel} & Sokoto        & 0.634            & 0.656   & 0.752 & 0.804 & 0.369 & 0.398 & 0.538 & 0.619 \\
            &Maiduguri     & 0.645            & 0.612   & 0.644 & 0.753 & 0.424 & 0.371 & 0.417 & 0.542 \\
            &Kano          & 0.681            & 0.663   & 0.760 & 0.916 & 0.488 & 0.481 & 0.614 & 0.809 \\
            &Katsina       & 0.667            & 0.624   & 0.661 & 0.748 & 0.507 & 0.421 & 0.445 & 0.537 \\\hline
\multirow{3}{*}{Midland}  &Kaduna        & 0.758            & 0.737   & 0.842 & 0.888 & 0.526 & 0.498 & 0.689 & 0.755 \\
            &Minna         & 0.662            & 0.697   & 0.802 & 0.781 & 0.423 & 0.516 & 0.627 & 0.610 \\
            &Yola          & 0.655            & 0.567   & 0.636 & 0.556 & 0.408 & 0.300 & 0.387 & 0.294 \\ \hline
\multirow{3}{*}{Guinea Savannah }  &Lokoja        & 0.727            & 0.763   & 0.871 & 0.897 & 0.586 & 0.629 & 0.731 & 0.769 \\
&Markudi       & 0.632            & 0.746   & 0.886 & 0.941 & 0.470 & 0.626 & 0.767 & 0.851 \\
&Warri         & 0.901            & 0.955   & 0.978 & 0.964 & 0.868 & 0.913 & 0.918 & 0.895 \\ \hline
\multirow{8}{*}{Coastal } &Lagos         & 0.850            & 0.930   & 0.960 & 0.980 & 0.808 & 0.861 & 0.892 & 0.913 \\
&Akure         & 0.800            & 0.852   & 0.889 & 0.874 & 0.740 & 0.718 & 0.768 & 0.736 \\
&Port Harcourt & 0.914            & 0.927   & 0.951 & 0.976 & 0.869 & 0.824 & 0.874 & 0.909 \\
&Owerri        & 0.883            & 0.937   & 0.952 & 0.960 & 0.837 & 0.855 & 0.882 & 0.883 \\
&Enugu         & 0.742            & 0.877   & 0.968 & 0.976 & 0.634 & 0.768 & 0.887 & 0.908 \\
&Calabar       & 0.921            & 1.000   & 1.000 & 1.000 & 0.893 & 0.939 & 0.949 & 0.948 \\
&Ogoja         & 0.755            & 0.861   & 0.953 & 0.966 & 0.676 & 0.756 & 0.966 & 0.901 \\
&Abeokuta      & 0.788            & 0.854   & 0.887 & 0.880 & 0.685 & 0.721 & 0.741 & 0.739\\ \hline
\end{tabular}
\end{table}

\begin{table}[]
\centering
  \caption{Trend analysis for SPI and SPEI for different locations at different time scales. Nonsignificant values at 95\% confidence interval are represented by *.}\label{trend}
  \footnotesize
\begin{tabular}{|l|l|llll|llll|}
                &                    &  \multicolumn{4}{c}{SPI}     &        \multicolumn{4}{c}{SPEI}          \\ \hline
                &                    &  \multicolumn{4}{c}{Scale}   &         \multicolumn{4}{c}{Scale}         \\ \hline
Region          & Location           & 1     & 3     & 6     & 12    & 1        & 3        & 6        & 12       \\ \hline
\multirow{4}{*}{Sahel} & Sokoto        & -0.00159 & -0.00214 & -0.00255* & -0.00349* & -0.00047 & -0.00071 & -0.00082 & -0.00135 \\
                & Maiduguri     & 0.0004   & 0.00107  & 0.00207  & 0.00354  & 0.00288  & 0.00373  & 0.00431  & 0.00645  \\
                & Kano          & 0.00132  & 0.00225  & 0.0037   & 0.00587  & 0.00403  & 0.00444  & 0.00474  & 0.007    \\
                & Katsina       & 0.00142  & 0.0024   & 0.00345  & 0.00457  & 0.00188  & 0.00273  & 0.00294  & 0.00356  \\ \hline
\multirow{3}{*}{Midland} & Kaduna        & -0.00026 & -0.00057 & -0.00069 & -0.00095 & -0.00047 & -0.00071 & -0.00082 & -0.00135 \\
                & Minna         & 0.00082  & 0.00131  & 0.00222  & 0.00362*  & 0.00124  & 0.00192  & 0.00325  & 0.00517  \\
                & Yola          & -0.00094 & -0.00099 & -0.00142* & -0.00201* & -0.00024 & -0.00001 & 0.00008  & -0.00013 \\ \hline
\multirow{3}{*}{Guinea Sav. } & Lokoja        & 0.00169  & 0.00249  & 0.00308  & 0.00369  & 0.00092  & 0.00176  & 0.00275  & 0.00377  \\
                & Markudi       & -0.00005 & 0.00024  & 0.00075  & 0.00082  & 0.00014  & 0.00055  & 0.00108  & 0.00095  \\
                & Warri         & -0.00002 & -0.00058 & -0.00094 & -0.00148 & 0.00007  & -0.00039 & -0.00075 & -0.00099 \\ \hline
\multirow{8}{*}{Coastal}    & Lagos         & -0.00008 & -0.00005 & -0.0002  & -0.00069 & -0.00001 & 0.00022  & 0.00009  & -0.00026 \\
                & Akure         & -0.00079 & -0.00131 & -0.0017  & -0.00211 & 0.00011  & 0.00004  & -0.00041 & -0.0006  \\
                & Port H. & 0.00016  & 0.00037  & 0.00024  & 0.00043  & 0.00025  & 0.00063  & 0.00067  & 0.0009   \\
                & Owerri        & -0.00044 & -0.00062 & -0.00082 & -0.00116 & -0.00003 & 0.00025  & 0.00015  & 0.0001   \\
                & Enugu         & 0.00047  & 0.00112  & 0.00116  & 0.00129  & 0.0006   & 0.00133  & 0.0017   & 0.00203  \\
                & Calabar       & 0.00063  & 0.00137  & 0.00149  & 0.00172  & 0.00063  & 0.00165  & 0.00183  & 0.00214  \\
                & Ogoja         & 0.00097  & 0.00174  & 0.00227  & 0.00254  & 0.00079  & 0.00146  & 0.00215  & 0.00286  \\
                & Abeokuta      & 0.00021  & 0.00068  & 0.00126  & 0.00171  & 0.00089  & 0.0016   & 0.00242  & 0.00325 \\ \hline
\end{tabular}
\end{table}

\begin{table}[]
\centering
  \caption{Frequency distribution of SPI and SPEI drought indices at 1-month timescale. EW = extremely wet, VW = very wet, MW = moderately wet, NN = near normal, MD = moderately dry, SD = severely dry, and ED = extremely dry.}\label{stats1}
  \footnotesize
\begin{tabular}{|l|l|lllllll|lllllll|}
\hline
Region          & Location      &\multicolumn{7}{c}{SPI}               & \multicolumn{7}{c}{SPEI}              \\ \hline
                &               & EW & VW & MW & NN & MD & SD & ED & EW & VW & MW & NN & MD & SD & ED \\ \hline
\multirow{4}{*}{Sahel}           & Sokoto        & 5             & 17       & 43             & 235         & 49             & 18           & 5             & 5             & 42       & 21             & 173         & 23             & 10           & 5             \\
                & Maiduguri     & 5             & 19       & 42             & 242         & 40             & 20           & 4             & 5             & 6        & 54             & 158         & 16             & 5            & 4             \\
                & Kano          & 4             & 21       & 40             & 250         & 35             & 16           & 6             & 5             & 8        & 52             & 159         & 10             & 11           & 3             \\
                & Katsina       & 3             & 20       & 42             & 252         & 28             & 21           & 6             & 5             & 13       & 21             & 185         & 14             & 8            & 2             \\ \hline
\multirow{3}{*}{Midland}    & Kaduna        & 9             & 15       & 38             & 249         & 35             & 21           & 5             & 5             & 42       & 21             & 173         & 23             & 10           & 5             \\
                & Minna         & 6             & 15       & 45             & 245         & 37             & 20           & 4             & 7             & 46       & 53             & 201         & 17             & 12           & 5             \\
                & Yola          & 8             & 15       & 31             & 255         & 38             & 20           & 5             & 3             & 40       & 24             & 176         & 19             & 12           & 5             \\ \hline
\multirow{3}{*}{Guinea} & Lokoja        & 7             & 15       & 43             & 243         & 43             & 15           & 6             & 10            & 9        & 37             & 245         & 26             & 8            & 6             \\
                & Markudi       & 7             & 17       & 37             & 251         & 39             & 13           & 8             & 7             & 20       & 59             & 253         & 17             & 13           & 3             \\
                & Warri         & 7             & 22       & 33             & 250         & 34             & 23           & 3             & 6             & 16       & 34             & 269         & 21             & 20           & 6             \\ \hline
\multirow{8}{*}{Coastal }    & Lagos         & 9             & 16       & 38             & 247         & 45             & 13           & 4             & 8             & 12       & 37             & 275         & 21             & 11           & 8             \\
                & Akure         & 7             & 18       & 36             & 246         & 38             & 22           & 5             & 3             & 14       & 43             & 269         & 17             & 22           & 4             \\
                & PH & 7             & 16       & 41             & 244         & 41             & 19           & 4             & 5             & 17       & 42             & 255         & 33             & 13           & 7             \\
                & Owerri        & 5             & 22       & 38             & 249         & 33             & 19           & 6             & 5             & 19       & 36             & 270         & 24             & 7            & 11            \\
                & Enugu         & 7             & 17       & 41             & 245         & 38             & 16           & 8             & 4             & 23       & 34             & 276         & 17             & 13           & 5             \\
                & Calabar       & 4             & 24       & 45             & 235         & 45             & 14           & 5             & 3             & 15       & 45             & 268         & 23             & 8            & 10            \\
                & Ogoja         & 10            & 18       & 28             & 262         & 31             & 16           & 7             & 9             & 18       & 32             & 273         & 24             & 9            & 7             \\
                & Abeokuta      & 5             & 23       & 35             & 239         & 37             & 17           & 4             & 7             & 16       & 30             & 275         & 14             & 10           & 8\\ \hline
\end{tabular}
\end{table}

\begin{table}[]
\centering
  \caption{Frequency distribution of SPI and SPEI drought indices at 3-month timescale. EW = extremely wet, VW = very wet, MW = moderately wet, NN = near normal, MD = moderately dry, SD = severely dry, and ED = extremely dry.}\label{stats3}
  \footnotesize
\begin{tabular}{|l|l|lllllll|lllllll|}
\hline
Region          & Location      &\multicolumn{7}{c}{SPI}               & \multicolumn{7}{c}{SPEI}              \\ \hline
                &               & EW & VW & MW & NN & MD & SD & ED & EW & VW & MW & NN & MD & SD & ED \\ \hline
\multirow{4}{*}{Sahel}          & Sokoto        & 4             & 18       & 45             & 221         & 50             & 19           & 3             & 7             & 40       & 25             & 211         & 27             & 12           & 8             \\
                & Maiduguri     & 4             & 21       & 34             & 230         & 52             & 14           & 5             & 8             & 7        & 54             & 197         & 19             & 10           & 5             \\
                & Kano          & 12            & 11       & 30             & 255         & 28             & 10           & 14            & 9             & 8        & 55             & 197         & 17             & 11           & 3             \\
                & Katsina       & 5             & 19       & 39             & 233         & 40             & 14           & 10            & 5             & 14       & 30             & 216         & 21             & 10           & 4             \\\hline
\multirow{3}{*}{Midland}    & Kaduna        & 5             & 14       & 37             & 240         & 37             & 22           & 5             & 7             & 40       & 25             & 211         & 27             & 12           & 8             \\
                & Minna         & 2             & 23       & 46             & 231         & 37             & 13           & 8             & 5             & 18       & 33             & 264         & 22             & 8            & 10            \\
                & Yola          & 6             & 13       & 43             & 233         & 39             & 21           & 5             & 7             & 38       & 61             & 213         & 23             & 9            & 9             \\\hline
\multirow{3}{*}{Guinea Sav. } & Lokoja        & 3             & 18       & 48             & 228         & 40             & 18           & 5             & 7             & 15       & 41             & 254         & 24             & 12           & 7             \\
                & Markudi       & 7             & 11       & 49             & 228         & 40             & 22           & 3             & 8             & 11       & 42             & 254         & 28             & 16           & 1             \\
                & Warri         & 7             & 19       & 31             & 244         & 36             & 18           & 5             & 6             & 18       & 32             & 244         & 34             & 19           & 7             \\\hline
\multirow{8}{*}{Coastal }   & Lagos         & 5             & 19       & 42             & 230         & 44             & 16           & 4             & 3             & 17       & 37             & 244         & 33             & 15           & 11            \\
                & Akure         & 7             & 18       & 35             & 233         & 45             & 17           & 5             & 2             & 14       & 42             & 235         & 35             & 23           & 9             \\
                & PH & 4             & 22       & 28             & 244         & 39             & 17           & 6             & 7             & 11       & 37             & 252         & 25             & 18           & 10            \\
                & Owerri        & 6             & 18       & 33             & 236         & 42             & 21           & 4             & 7             & 12       & 30             & 249         & 37             & 18           & 7             \\
                & Enugu         & 7             & 14       & 31             & 241         & 46             & 14           & 7             & 2             & 18       & 40             & 257         & 20             & 12           & 11            \\
                & Calabar       & 7             & 17       & 36             & 239         & 39             & 12           & 10            & 8             & 9        & 40             & 250         & 26             & 16           & 11            \\
                & Ogoja         & 5             & 21       & 31             & 248         & 30             & 20           & 5             & 8             & 19       & 28             & 260         & 23             & 12           & 10            \\
                & Abeokuta      & 6             & 19       & 33             & 232         & 38             & 15           & 5             & 7             & 9        & 44             & 232         & 30             & 18           & 8\\ \hline
\end{tabular}
\end{table}

\begin{table}[]
\centering
  \caption{Frequency distribution of SPI and SPEI drought indices at 6-month timescale. EW = extremely wet, VW = very wet, MW = moderately wet, NN = near normal, MD = moderately dry, SD = severely dry, and ED = extremely dry.}\label{stats6}
  \footnotesize
\begin{tabular}{|l|l|lllllll|lllllll|}
\hline
Region          & Location      &\multicolumn{7}{c}{SPI}               & \multicolumn{7}{c}{SPEI}              \\ \hline
                &               & EW & VW & MW & NN & MD & SD & ED & EW & VW & MW & NN & MD & SD & ED \\ \hline
\multirow{4}{*}{Sahel}           & Sokoto        & 4             & 15       & 38             & 238         & 41             & 22           & 2             & 7             & 12       & 26             & 254         & 33             & 18           & 10            \\
                & Maiduguri     & 3             & 20       & 34             & 228         & 51             & 22           & 2             & 9             & 15       & 27             & 258         & 29             & 14           & 8             \\
                & Kano          & 13            & 11       & 25             & 254         & 35             & 11           & 11            & 10            & 12       & 36             & 250         & 31             & 18           & 3             \\
                & Katsina       & 0             & 23       & 46             & 225         & 43             & 20           & 3             & 6             & 17       & 42             & 248         & 26             & 17           & 4             \\\hline
\multirow{3}{*}{Midland}    & Kaduna        & 6             & 12       & 39             & 240         & 34             & 26           & 3             & 7             & 12       & 26             & 254         & 33             & 18           & 10            \\
                & Minna         & 1             & 20       & 51             & 225         & 40             & 16           & 7             & 2             & 15       & 36             & 258         & 21             & 11           & 17            \\
                & Yola          & 7             & 15       & 31             & 244         & 33             & 23           & 7             & 4             & 18       & 25             & 263         & 27             & 9            & 14            \\\hline
\multirow{3}{*}{Guinea Sav. } & Lokoja        & 3             & 24       & 39             & 231         & 40             & 19           & 4             & 9             & 17       & 31             & 246         & 33             & 18           & 6             \\
                & Markudi       & 7             & 17       & 27             & 249         & 31             & 24           & 5             & 9             & 15       & 28             & 245         & 34             & 28           & 1             \\
                & Warri         & 4             & 17       & 44             & 239         & 31             & 22           & 3             & 12            & 11       & 38             & 243         & 39             & 12           & 5             \\\hline
\multirow{8}{*}{Coastal }    & Lagos         & 2             & 22       & 36             & 237         & 35             & 24           & 4             & 0             & 21       & 34             & 248         & 27             & 16           & 14            \\
                & Akure         & 9             & 15       & 30             & 242         & 39             & 20           & 5             & 5             & 14       & 31             & 247         & 31             & 21           & 11            \\
                & PH & 4             & 21       & 33             & 239         & 38             & 21           & 4             & 5             & 20       & 36             & 239         & 35             & 18           & 7             \\
                & Owerri        & 8             & 12       & 37             & 244         & 27             & 28           & 4             & 10            & 12       & 33             & 244         & 32             & 19           & 10            \\
                & Enugu         & 5             & 12       & 38             & 233         & 51             & 14           & 7             & 4             & 9        & 41             & 258         & 26             & 7            & 15            \\
                & Calabar       & 7             & 18       & 29             & 252         & 30             & 16           & 8             & 10            & 16       & 25             & 261         & 27             & 10           & 11            \\
                & Ogoja         & 4             & 22       & 39             & 237         & 32             & 22           & 4             & 11            & 18       & 33             & 244         & 30             & 14           & 10            \\
                & Abeokuta      & 2             & 20       & 42             & 229         & 34             & 16           & 5             & 6             & 11       & 40             & 228         & 30             & 28           & 5 \\ \hline
\end{tabular}
\end{table}

\begin{table}[]
\centering
  \caption{Frequency distribution of SPI and SPEI drought indices at 12-month timescale. EW = extremely wet, VW = very wet, MW = moderately wet, NN = near normal, MD = moderately dry, SD = severely dry, and ED = extremely dry.}\label{stats12}
  \footnotesize
\begin{tabular}{|l|l|lllllll|lllllll|}
\hline
Region          & Location      &\multicolumn{7}{c}{SPI}               & \multicolumn{7}{c}{SPEI}              \\ \hline
                &               & EW & VW & MW & NN & MD & SD & ED & EW & VW & MW & NN & MD & SD & ED \\ \hline
\multirow{4}{*}{Sahel}           & Sokoto        & 3             & 15       & 48             & 215         & 57             & 22           & 0             & 7             & 12       & 15             & 262         & 24             & 29           & 11            \\
                & Maiduguri     & 1             & 20       & 43             & 228         & 47             & 20           & 1             & 0             & 29       & 22             & 246         & 29             & 22           & 12            \\
                & Kano          & 7             & 26       & 20             & 253         & 40             & 2            & 12            & 1             & 30       & 34             & 222         & 42             & 31           & 0             \\
                & Katsina       & 3             & 17       & 39             & 227         & 51             & 23           & 0             & 2             & 16       & 66             & 216         & 28             & 25           & 7             \\ \hline
\multirow{3}{*}{Midland}    & Kaduna        & 5             & 15       & 36             & 235         & 36             & 29           & 4             & 7             & 12       & 15             & 262         & 24             & 29           & 11            \\
                & Minna         & 7             & 12       & 36             & 248         & 27             & 21           & 9             & 0             & 14       & 40             & 263         & 7              & 11           & 25            \\
                & Yola          & 5             & 23       & 21             & 245         & 35             & 27           & 4             & 0             & 13       & 42             & 255         & 24             & 5            & 21            \\ \hline
\multirow{3}{*}{Guinea Sav. } & Lokoja        & 2             & 25       & 38             & 233         & 33             & 29           & 0             & 8             & 14       & 48             & 223         & 42             & 17           & 8             \\
                & Markudi       & 12            & 5        & 41             & 242         & 37             & 17           & 6             & 15            & 19       & 13             & 252         & 35             & 20           & 6             \\
                & Warri         & 3             & 21       & 47             & 224         & 43             & 19           & 3             & 10            & 12       & 37             & 238         & 44             & 14           & 5             \\ \hline
\multirow{8}{*}{Coastal }    & Lagos         & 6             & 16       & 42             & 235         & 37             & 21           & 3             & 3             & 19       & 38             & 247         & 27             & 13           & 13            \\
                & Akure         & 12            & 12       & 24             & 247         & 40             & 21           & 4             & 17            & 7        & 24             & 256         & 23             & 22           & 11            \\
                & PH & 10            & 10       & 39             & 241         & 37             & 18           & 5             & 3             & 17       & 51             & 228         & 33             & 22           & 6             \\
                & Owerri        & 1             & 29       & 28             & 237         & 39             & 26           & 0             & 12            & 13       & 23             & 262         & 22             & 13           & 15            \\
                & Enugu         & 5             & 15       & 32             & 239         & 51             & 12           & 6             & 0             & 14       & 49             & 250         & 31             & 3            & 13            \\
                & Calabar       & 3             & 22       & 37             & 233         & 45             & 16           & 4             & 20            & 14       & 26             & 242         & 39             & 17           & 2             \\
                & Ogoja         & 1             & 25       & 56             & 221         & 45             & 6            & 6             & 11            & 20       & 48             & 226         & 39             & 10           & 6             \\
                & Abeokuta      & 0             & 22       & 56             & 211         & 40             & 17           & 2             & 3             & 17       & 46             & 217         & 33             & 20           & 12 \\ \hline
\end{tabular}
\end{table}

\clearpage
\newpage

\end{document}